\documentclass[sigconf]{acmart}
\AtBeginDocument{%
  }
\usepackage{array}
\usepackage{tabularx}
\usepackage{booktabs}
\usepackage{multirow}
\usepackage{hyperref}
\usepackage[table]{xcolor}
\usepackage[most]{tcolorbox}
\usepackage{orcidlink}
\usepackage[export]{adjustbox}
\usepackage{float}
\usepackage{subcaption}
\usepackage{tikz}
\usetikzlibrary{arrows.meta, positioning, shapes.geometric}

\tcbset{
  conclusionbox/.style={
    colback=gray!15,
    colframe=black,
    boxrule=0.8pt,
    arc=2pt,
    left=8pt,
    right=8pt,
    top=6pt,
    bottom=6pt,
    before skip=6pt,
    after skip=10pt,
    halign=center 
  }
}

\setcopyright{acmlicensed}
\copyrightyear{2026}
\acmYear{2026}
\acmDOI{XXXXXXX.XXXXXXX}
\acmConference[EASE 2026]{The 30th International Conference on Evaluation and Assessment in Software Engineering}{9–12 June, 2026}{Glasgow, Scotland, United Kingdom}
\acmISBN{978-1-4503-XXXX-X/2018/06}

\begin{document}

\title[How Do Software Engineering Students Use Generative AI in Real-World Capstone Projects?]{How Do Software Engineering Students Use Generative AI in Real-World Capstone Projects? An Empirical Baseline Study}

\author{Michael Mircea, Elisa Schmid, Jakob Droste, Kurt Schneider}
\affiliation{%
  \institution{Leibniz University Hannover\\Software Engineering Group}
  \city{Hannover}
  \country{Germany}
}
\email{{michael.mircea, elisa.schmid, jakob.droste, kurt.schneider}@inf.uni-hannover.de}
\orcid{0009-0007-6783-4981}

\renewcommand{\shortauthors}{Mircea et al.}

\begin{abstract}
Real-world Capstone Projects (RWCPs) are a key component of software engineering education, enabling students to develop software for external clients under authentic conditions. Their high ecological validity, combined with substantial variation in domains, technologies, and stakeholders, typically requires flexible and minimally prescriptive teaching approaches. The rapid integration of generative AI (GenAI) into professional software development adds new challenges: students are expected to use AI tools that are common in practice, yet unguided use may affect learning, collaboration, and consistency in ways that are not yet well understood.

To establish an empirical baseline for responsible GenAI integration, we conducted a large-scale study of self-determined GenAI use in an undergraduate RWCP course. The module involved 178 students working in 18 teams across 15 client projects over four months, with GenAI use explicitly permitted. We collected mixed-method survey data from 150 students on attitudes, usage prevalence, workflows, use cases, and perceived benefits and risks, and surveyed client stakeholders regarding expectations and concerns.

Our findings provide (1) a characterization of GenAI practices across the software engineering lifecycle, including a distinction between emerging workflows; (2) student-recommended use cases and responsible-use directives emphasizing verification and maintaining independent understanding; (3) client perspectives highlighting strong support for GenAI use but clear expectations regarding understanding, quality, and data protection; and (4) implications for future course iterations, including the need for explicit responsible-use guidelines, targeted AI literacy resources, and team-level governance roles. This study offers a status quo baseline for evidence-based pedagogical interventions in the era of GenAI.
\end{abstract}



\received{02.03.2026}
\received[revised]{27.04.2026}
\received[accepted]{30.04.2026}

\maketitle

\section{Introduction}

Real-world capstone projects (RWCPs) provide software engineering (SE) students with authentic experience in client collaboration, evolving requirements, and team-based software development~\cite{bastarrica2017whatcanstudentsget}. Their high ecological validity, combined with substantial variation in domains, technologies, and stakeholder expectations, typically necessitates flexible and minimally prescriptive teaching approaches~\cite{tenhunen2023systematic}. This flexibility becomes particularly relevant with the rapid rise of generative AI (GenAI), which is increasingly embedded in professional SE practice~\cite{simaremare2024state}.

While allowing students to use GenAI freely in RWCPs preserves authenticity and mirrors current industry behavior~\cite{nanda2025state}, the educational consequences of such unguided use remain unclear. Emerging literature on GenAI in SE education has begun to explore its potential benefits and risks, but many studies focus on controlled settings, single tasks, or individual programming assignments~\cite{denny2024computing,daun2023chatgpt,choudhuri2024howfar,choudhuri2025insights}. Little is known about how students adopt and integrate GenAI in complex, multi-month, results-oriented projects, or how this aligns with client expectations in such authentic environments. The absence of baseline evidence limits the development of informed instructional strategies, governance mechanisms, and targeted support for effective GenAI use in capstone contexts.

To address this research gap, we conducted a large-scale baseline study of self-determined GenAI use in an undergraduate RWCP course involving 178 students, 18 teams, and 15 client-sourced projects. GenAI use was explicitly permitted throughout the four-month module, provided students adhered to data-sensitivity guidelines. We collected mixed-method survey data from 150 students capturing attitudes toward GenAI, actual usage patterns across the SE lifecycle, perceived benefits and risks, and recommended practices. A complementary client survey elicited expectations, concerns, and perspectives on student GenAI use.

This paper contributes: (1) an empirical characterization of student GenAI use in authentic SE capstones, including attitudes, workflows, and use cases; (2) a qualitative account of perceived strengths, risks, and responsible-use practices; (3) an analysis of client perspectives on GenAI in RWCPs; and (4) implications for teaching, assessment, and future iterations of capstone governance in the era of GenAI.

\section{Background}
\subsection{Capstone Projects in SE Education}

Capstone courses are a common component of SE programs, intended to provide students with authentic, practice-oriented project experience. Tenhunen et al.'s recent systematic review of 127 studies \cite{tenhunen2023systematic} shows that SE capstones typically involve team-based development of a substantial software system, often for an external client. Most courses run for one semester, rely on teams of four to five students, and require students to engage in core SE activities such as requirements elicitation, design, implementation, and testing. 58~\% of reviewed capstones include real-world projects, with clients providing requirements, feedback, and domain knowledge. These authentic settings emphasize both technical and professional skills—particularly teamwork, communication, and project planning—which are consistently reported as central learning outcomes. However, they also provide highly varying conditions and constraints resulting in unusually high autonomy for students. This makes generalizable teaching methods or technology restrictions impractical. Assessment practices vary widely but commonly combine artifact evaluation, process assessment, peer evaluation, and client feedback.  

\subsection{GenAI in SE Education}
The rapid adoption of GenAI tools such as ChatGPT and GitHub Copilot has prompted growing interest in their role in software engineering SE education~\cite{denny2024computing}. Due to the technology only recently becoming widely accessible, many works are based on positions or conceptual in nature~\cite{daun2023chatgpt,denny2024computing} and highlight transformative potential, opportunities, and challenges.

Early, small-scale empirical work has examined how students use GenAI to support both learning and programming tasks, identifying benefits such as clarification of concepts, help with syntax, and support for initial implementation steps, as well as challenges including misleading explanations, low-quality code, and difficulty in more complex implementation settings \cite{choudhuri2025insights}. Furthermore, free conversational GenAI use by students may not lead to statistically significant productivity increases in SE tasks while increasing frustration levels~\cite{choudhuri2024howfar}. These findings raise further questions about unguided GenAI use, particularly in more complex SE implementation tasks such as RWCPs, and how students can be actively assisted.

\subsection{GenAI in SE Capstones}
Some existing works evaluated the integration of AI tools into SE capstone projects~\cite{roy2025empowering,gonzalez2022improving,neyem2024exploring}. For example, Gonzalez et al.~\cite{gonzalez2022improving} evaluated the effectiveness of virtual assistants in capstones, while Neyem et al.~\cite{neyem2024exploring} explored the potential of GenAI to assist in standup report recommendations in capstones. While these active interventions present reasonable use cases and often yield promising results, it is unclear how they perform against a self-determined GenAI utilization baseline. The absence of such a baseline could also obscure further use cases in which an active intervention would benefit students the most when using GenAI.

Overall, existing work provides first insights into GenAI use for isolated SE tasks and explores targeted interventions within capstone settings, the literature lacks an understanding of how students naturally appropriate GenAI in authentic project environments. No prior study, to our knowledge, has broadly examined autonomous GenAI use across multiple RWCPs, nor compared student behavior with client expectations. This gap limits the design of evidence-based teaching guidelines and responsible-use policies for GenAI in highly autonomous project contexts. Our work addresses this issue by providing a large-scale baseline characterization of self-determined GenAI use in RWCPs during a moment in time where such tools are widely adopted in the general population, capturing both student and client perspectives to inform future pedagogical and technological interventions.

\subsection{Course Description: The SWP}

The software engineering capstone project at Leibniz University Hannover, referred to as the \textit{Software Project} (SWP), has been conducted annually for over 20~years. The 2025/26 iteration involved 178 students forming 18 teams that developed 15 real-world, client-sourced software projects. The teaching team consisted of six tutors, each supervising three teams, and two instructors, plus the responsible professor overseeing course design, assessment, and client coordination. This description follows the reporting structure recommended by Tenhunen et al.~\cite{tenhunen2023systematic}.

\subsubsection{Course Structure and Placement}
The SWP is a mandatory four-month capstone in the final year of the computer science undergraduate program. It provides twice the credit volume of regular courses and requires approximately 15~hours of weekly effort. Learning objectives span technical, socio-technical, and organizational competencies, including programming, collaboration and communication, requirements engineering, stakeholder interaction, project management, quality assurance, and product presentation.

The course is organized into four blocks: \emph{Exploration}, \emph{First Iteration}, \emph{Second Iteration}, and \emph{Polishing}. It follows a hybrid development structure combining traditional and agile practices. Expected deliverables include: (1) a client-signed Software Requirements Specification (SRS) based on the Volere template~\cite{robertson2000volere}, containing use cases, architectural considerations, and acceptance criteria; (2) feature-complete software increments at the end of each iteration, with a mid-course demo to all teams; and (3) a final software system with accompanying documentation, installation guide, and a poster for a public presentation at the course conclusion.

\subsubsection{Student Teams}
An unexpected 20\% enrollment increase resulted in larger-than-usual teams of 9--10 students (typically 7--8). Teams were algorithmically assembled, highly prioritizing students' project preferences and secondarily balancing self-reported technical and soft-skill levels. All students contributed to development work, complemented by non-technical tasks depending on project needs. Each team autonomously assigned three internal roles: team lead, roll-out manager, and quality assurance lead. A dedicated tutor supported each team and participated in its weekly client meetings.

\subsubsection{Clients and Projects}
15 client organizations provided project proposals, each represented by 2--4 stakeholders. All projects were authentic real-world requests with operational or prototyping value. Client domains were highly diverse, including IT, automotive, scientific research, health, logistics, and public administration. Consequently, project technologies ranged from command-line tools to full-stack web applications and cross-platform systems. Notably, five of the fifteen projects contained GenAI as a program component of the project itself this iteration, mirroring the increasing prevalence of GenAI applications in industry.

To maintain comparability across projects despite this heterogeneity, instructors pre-negotiated minimum viable product (MVP) requirements with clients prior to course start, while leaving room for optional extensions for high-performing teams. Due to student demand, three client projects were offered twice, resulting in 18 projects in total. Clients were required to attend weekly meetings and often engaged with teams beyond these sessions through additional meetings, asynchronous communication, or on-site visits.

\subsubsection{Process and Infrastructure}
While the SWP emphasizes team autonomy, a lightweight process structure ensures progress and supports learning outcomes. In addition to the overarching four-block-structure, weekly meetings also followed a rigid structure and consisted of: (1) a short team-internal SCRUM (5-10 minutes), initially led by the tutor but then taken over by the team lead; (2) a client meeting (20-40 minutes) focused on requirements, progress, and prototype evaluation, informed by a mandatory agile board maintained throughout development; and (3) a rotating one-on-one conversation between tutor and a team member (5-15 minutes), used to surface conflicts and verify individual contributions through discussion of artifacts and git commit histories.

All teams gave a mid-course presentation summarizing their progress and an end-of-course poster presentation of their finished project. Due to technological diversity, infrastructural support was intentionally minimal: each team received a mandatory GitLab repository (including issue tracking and agile tooling) and optional access to a dedicated deployment server for testing and delivery.

GenAI use was explicitly permitted at students’ discretion without disclosure requirements. Students received mandatory guidance on responsible and data-sensitive GenAI use, especially regarding client data. The university also offered its own data-isolated ChatGPT instance (LUHKI), though adoption was low.

\subsubsection{Assessment and Evaluation}
Assessment in the SWP reflects the complexity of real-world capstones. Students typically pass if their project is successful and they demonstrate active participation. Tutors monitored individual engagement through commit activity, client interaction, and contributions to artifacts. Cases of insufficient participation, conflict, or project failure (rare) triggered an escalation to instructors, who may conduct individual evaluations.

Deliverables were assessed for timely completion and acceptable quality. The SRS underwent tutor review to ensure soundness of use cases and acceptance test cases before sign-off. During the final week, all acceptance criteria in the SRS were tested with the client to determine project success; criteria could be adapted during development following agile practice. In addition to external assessment, students completed a retrospective at the end of the course, which, while not graded, encouraged metacognitive learning.

\section{Methodology}
Our study was conducted at the end of the capstone project and aimed at capturing reflective feedback based in students' actual project experience. We combined closed and open survey questions to gather as much insight as possible into how students use and regard GenAI in the SWP, as well as what clients expect and are concerned about. Both surveys were conducted in German and can be found in full in our supplementary material~\cite{mircea_2026_18836366}.

\subsection{Student Survey}
The student survey was administered during the final regular session of the SWP and aimed to capture reflective feedback grounded in students’ actual project experience. Given the exploratory nature of our research goals, the survey combined structured closed questions with open-ended items to obtain both quantifiable patterns and deeper qualitative insights. To allow students to express both positive and negative perceptions, many items used semantic differential scales rather than unidirectional Likert scales. The survey consisted of three parts and was conducted in German.

\subsubsection{General Data}
The first part collected background information to contextualize later responses. Students answered three closed questions:

\begin{itemize}
    \item \textbf{[Q1.1]} \textit{How do you assess your own programming skills?} (``Low'', ``Middle'', ``High'')
    \item \textbf{[Q1.2]} \textit{What is your prior, professional software development experience?} (``None'' to ``Over 3~years'')
    \item \textbf{[Q1.3]} \textit{What is your general attitude towards GenAI (not limited to the SWP)?} (``Very pessimistic'' to ``Very optimistic'')
\end{itemize}

Question~\textbf{[Q1.3]} is particularly relevant as a baseline measure, since broader shifts in student perception and technological maturity may influence how students evaluate GenAI use or interventions in future cohorts.

\subsubsection{GenAI Usage Reporting}
The second part captured how students used GenAI throughout the project, focusing on usage frequency, use cases, and modes of interaction with GenAI tools:

\begin{itemize}
    \item \textbf{[Q2.1]} \textit{How often did you use GenAI for the SWP (across all purposes)?} (``Never'' to ``Constantly'')
    \item \textbf{[Q2.2]} \textit{For which purposes did you use GenAI during the SWP?} (multiple choice)
    \begin{itemize}
        \item Learning new technologies (e.g., frameworks, languages)
        \item Requirements elicitation or domain analysis
        \item Documentation (e.g., SRS support, code comments)
        \item High-level design (e.g., architecture suggestions)
        \item Generating or debugging code
        \item Problem solving (non-code)
        \item Qualitative improvement or testing of code
        \item Other (free text)
    \end{itemize}
    \item \textbf{[Q2.3]} \textit{Which types of GenAI tools did you use for the SWP?} (``Universal chatbots'', ``Integrated tools or specialized models'', ``LUHKI'', ``Other'')
    \item \textbf{[Q2.4]} \textit{How often did you generate solutions with GenAI and then check, improve, or adapt them?} (``Never'' to ``Very frequently'')
    \item \textbf{[Q2.5]} \textit{How often did you provide your own (preliminary) solutions and have GenAI check, improve, or adapt them?} (``Never'' to ``Very frequently'')
\end{itemize}

Questions~\textbf{[Q2.4]} and \textbf{[Q2.5]} were included to distinguish between \emph{AI-in-the-loop} (AITL) and \emph{human-in-the-loop} (HITL) workflows~\cite{natarajan2025human} commonly discussed in emerging GenAI literature.

\subsubsection{GenAI Assessment}
The final part addressed students’ evaluation of GenAI in the SWP and invited reflective feedback:

\begin{itemize}
    \item \textbf{[Q3.1]} \textit{Overall assessment of GenAI for the SWP} (``Very harmful'' to ``Very useful'')
    \item \textbf{[Q3.2]} \textit{Based on your experience, what would you recommend to future SWP students regarding GenAI use, and what would you advise against?} (optional free text)
    \item \textbf{[Q3.3]} \textit{Do you have any further feedback for the SWP organization regarding GenAI?} (optional free text)
\end{itemize}

Responses to \textbf{[Q3.3]} were sparse and did not yield findings relevant to this study, and are therefore not further analyzed.

\subsection{Client Survey}
The client survey was conducted after the student survey and aimed to capture clients’ general attitudes toward GenAI use in their projects as well as specific areas of concern. The survey began with a short disclaimer informing clients that a large majority of students in this cohort had used GenAI to some extent during the SWP. This disclosure served to surface any potential concerns or expectations related to this usage.

The survey was administered via email, participation was voluntary and anonymous, and it consisted of five questions:

\begin{itemize}
    \item \textbf{[Q4.1]} \textit{What is your general attitude towards GenAI (not limited to the SWP)?} (``Very pessimistic'' to ``Very optimistic'')
    \item \textbf{[Q4.2]} \textit{Did your project this year explicitly include GenAI as part of the project proposal?} (Yes/No)
    \item \textbf{[Q4.3]} \textit{How do you view the high student usage of GenAI for the development of your software project?} (``Critical'' to ``Harmless'')
    \item \textbf{[Q4.4]} \textit{Which concerns do you have regarding students’ diverse GenAI usage in this year's or in future SWP projects?}
    \begin{itemize}
        \item Functional correctness
        \item Insufficient quality
        \item Infringement of intellectual property
        \item Data protection risks
        \item Lack of transparency or accountability
        \item Students lacking own understanding in client meetings
        \item I have no concerns
        \item Other (please specify)
    \end{itemize}
    \item \textbf{[Q4.5]} \textit{If you were to participate as a client again next year, what would you prefer regarding students’ GenAI use?}
    \begin{itemize}
        \item Students should not use GenAI
        \item Students should use GenAI rarely and with care
        \item Students may use GenAI as they see fit
        \item Students should be actively supported in using GenAI through teaching content and tools
    \end{itemize}
\end{itemize}

\subsection{Ethics and Consent}
Students were given 15~minutes during the final regular session of the course to complete the survey, but participation was explicitly voluntary, anonymous, and had no impact on course assessment. The survey began with a disclaimer explaining its purpose, namely to support research on GenAI use in the SWP and to inform improvements for future iterations. Students were reminded that GenAI use in the SWP was fully permitted and were encouraged to respond honestly based on their experience.

Client stakeholders received a similar invitation via email, outlining that their participation was voluntary and anonymous and that responses would be used for research and alignment of future course design with client expectations and concerns.

\subsection{Analysis Methods}

\subsubsection{Quantitative Analysis}
We used two methods for the quantitative analysis of the responses from the student survey. First, we present the raw data from some demographic questions in graphical form, as these are predominantly descriptive data or frequency distributions of the methods and GenAI models used. Second, we tested several hypotheses, shown in \autoref{tab:null_hypotheses}, to systematically analyze correlations between selected factors such as self-assessed programming skills, GenAI usage frequency, attitudes towards GenAI, and perceived benefits. Because these variables are measured on ordinal scales and do not fulfill the assumptions of parametric tests, we use Spearman’s rank correlation~\cite{spearman1904association} for these analyses. 

\begin{table}[htbp]
    \centering
    \caption{Null hypotheses for the Spearman rank correlation analyses}
    \label{tab:null_hypotheses}
    \begin{tabular}{p{1cm} p{6.8cm}} 
        \toprule
        \textbf{$H_0$} & \textbf{Description} \\
        \midrule
        $H1_{0}$ & There is no relation between a student's \textit{self-assessed programming skills} and their \textit{frequency of GenAI usage} in the context of the SWP. \\
        \addlinespace
        $H2_{0}$ & There is no correlation between a student's \textit{frequency of GenAI usage} for a student project and their \textit{perceived benefits of GenAI} for the SWP.\\
        \addlinespace
        $H3_{0}$ & There is no relation between a student's \textit{attitude towards the use of GenAI} and their frequency of use of the \textit{GenAI-in-the-loop} approach. \\
        \addlinespace
        $H4_{0}$ & There is no relation between a student's \textit{attitude towards the use of GenAI} and their frequency of use of the \textit{human-in-the-loop} approach. \\
        \addlinespace
        $H5_{0}$ & There is no relation between a student's frequency of use of the \textit{GenAI-in-the-loop} approach and the \textit{human-in-the-loop} approach. \\
        \bottomrule
    \end{tabular}
\end{table}

\subsubsection{Qualitative Analysis}
We analyzed the qualitative feedback of the survey by means of a three-step coding procedure, following the guidelines of Saldaña~\cite{saldana2021coding}. The entire coding procedure was conducted by two experts (authors of this paper), both of whom have a research background in human-centered AI.

The first coding cycle was performed as \textit{In-Vivo} coding. First, the two raters independently scanned the entire data set to get an overview of the potential codes. Then they analyzed the statements together, one by one, until no new codes could be identified. Finally, this set of codes was refined into guidelines for \textit{pattern} coding.

The second coding cycle was performed as \textit{pattern} coding on the first half of the data set, to test the reliability of the previously developed guidelines. The two raters labeled 44 out of 89 statements and then checked for disagreements between each other. A superficial analysis revealed that the raters disagreed on roughly one-third of the statements. They finalized the coding guidelines accordingly, which can be found in our supplementary material~\cite{mircea_2026_18836366}.

The third and final coding cycle was performed as \textit{pattern} coding on the entire data set, using the updated guidelines and the qualitative data analysis software \textit{MaxQDA}. The inter-rater agreement was calculated using Brennan \& Prediger $\kappa$~\cite{brennan81kappa}, a chance-corrected variant of Cohen's $\kappa$. The final agreement was determined at $\kappa = 79\%$. In accordance with the guidelines of Landis and Koch~\cite{landis1977measurement}, this constitutes a \textit{substantial}, close to \textit{almost perfect} agreement between the raters. For the purpose of numerical analysis, both raters discussed and resolved all remaining conflicts. The results of the coding procedure can be found in our supplementary material~\cite{mircea_2026_18836366}.

\section{Results}
This section presents the results of the two surveys. 150 out of 178 students participated in the survey, with 89 of them providing optional, sometimes extensive, free-text responses. Out of 15 client organizations, eleven individuals completed the client survey. The full result data can be found in our supplementary material~\cite{mircea_2026_18836366}.

\subsection{Quantitative Student Survey Results}

Before presenting the qualitative analysis of the students’ free-text responses, we report the quantitative survey results. These comprise descriptive statistics as well as selected Spearman's rank correlation analyses~\cite{spearman1904association}. 139 out of 150 participating students reported that they used GenAI at least to some extent.

\begin{figure}[htbp]
    \centering
    \includegraphics[width=0.45\textwidth]{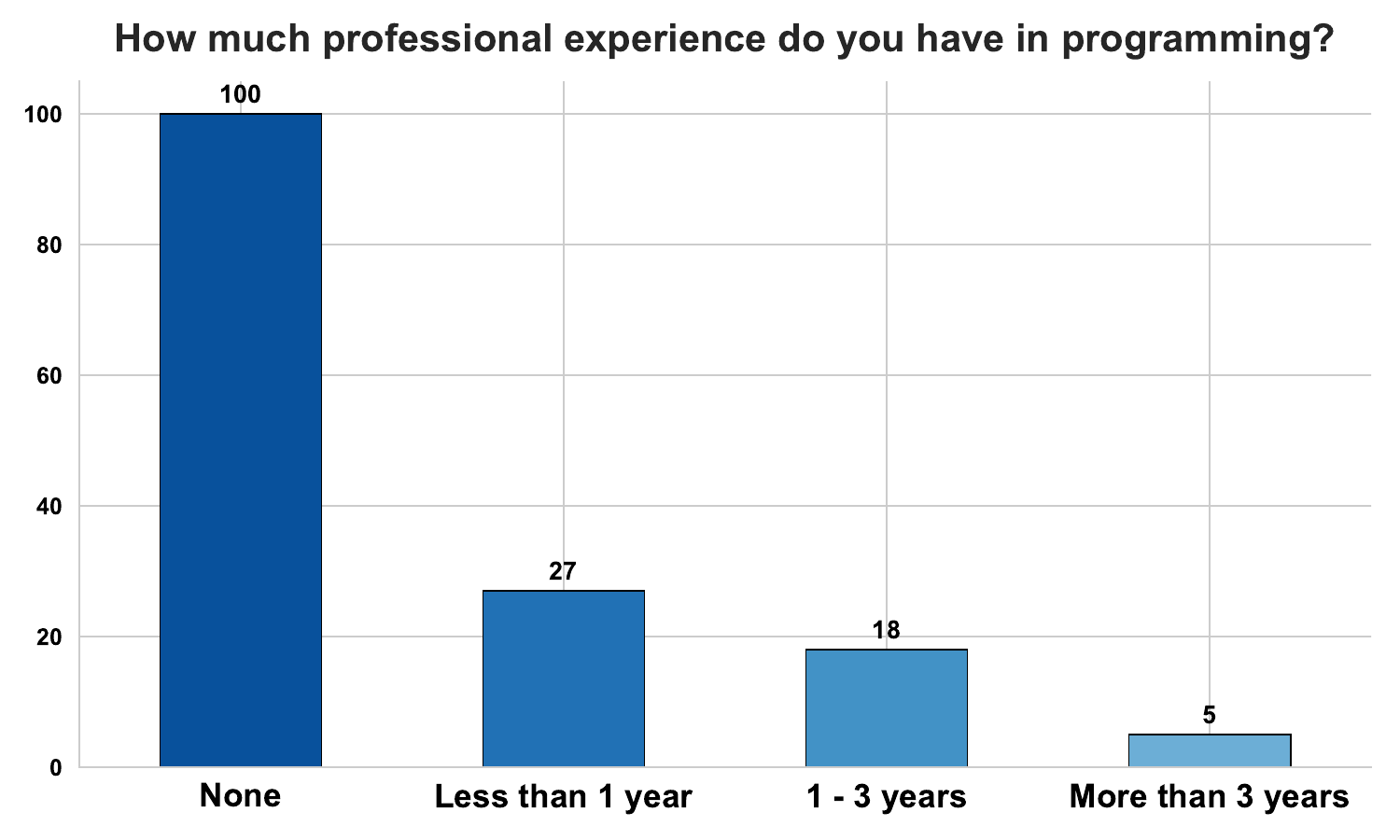}
    \caption{Programming experience of the 150 participants, divided into four levels of prior professional experience.}
    \label{fig:experience}
\end{figure}

\autoref{fig:experience} shows the distribution of professional programming experience among the 150 respondents. We distinguished four levels of experience: no professional experience, less than one year, one to three years, and more than three years. The majority of students (100) reported having no professional programming experience. Only one third indicated any prior professional experience.

\begin{figure}[htbp]
    \centering
    \includegraphics[width=0.45\textwidth]{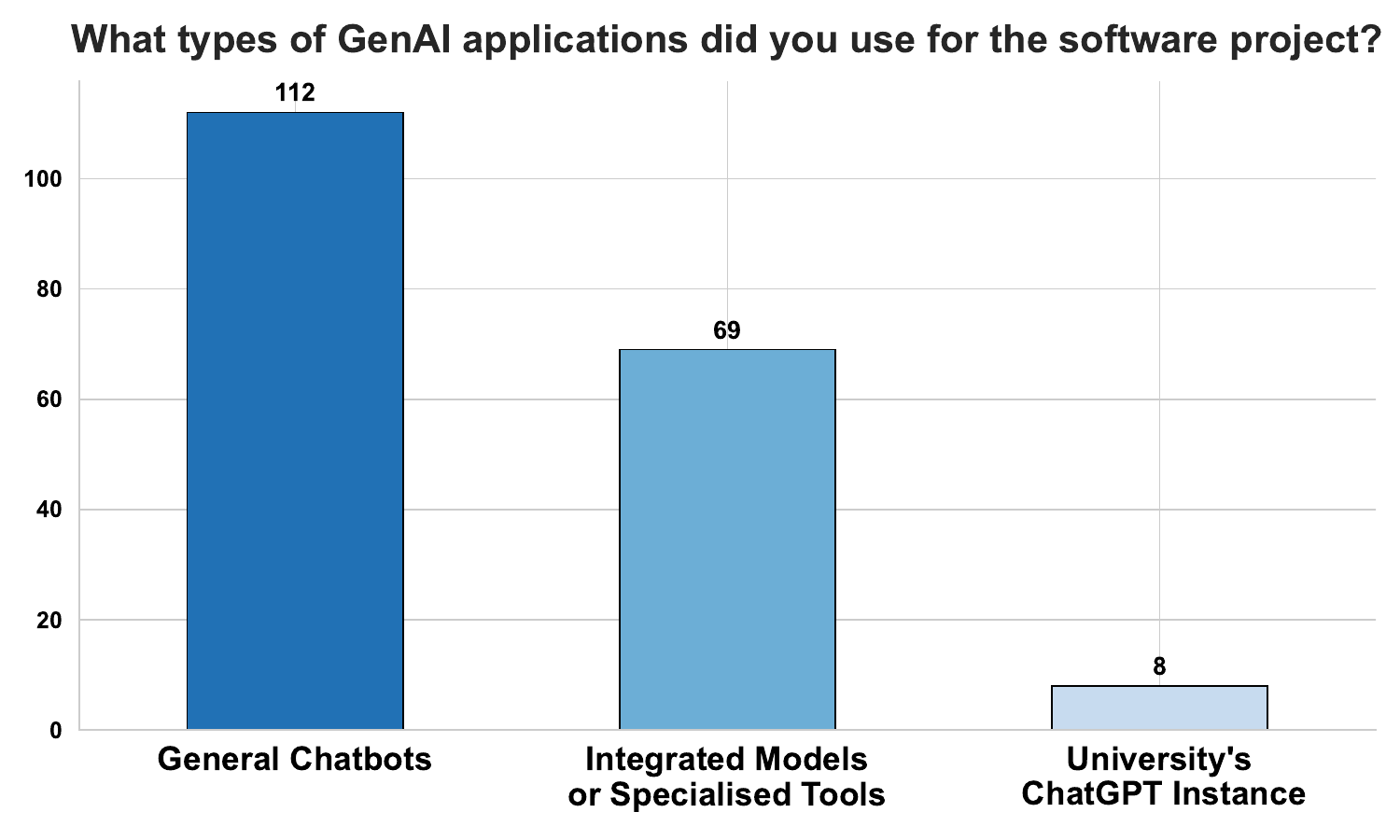}
    \caption{Distribution of GenAI application type use among 150 participants. Multiple selections were possible.}
    \label{fig:type}
\end{figure}

Students were asked which types of GenAI applications they used in their software project, if any. We distinguished three categories: general chatbots (e.g., ChatGPT, Gemini, Claude, Le Chat), integrated models or specialized tools (e.g., GitHub Copilot, Cursor), and LUHKI (the university’s own ChatGPT instance). Because these types serve different purposes, students could select multiple categories. The results are visualized in \autoref{fig:type}. If students used GenAI tools not listed among the response options, they could specify them in a free-text field. We received three such responses: one student stated they did not use GenAI at all, while two students indicated using LLM agents and self-hosted LLMs.

\begin{figure}[htbp]
    \centering
    \includegraphics[width=0.45\textwidth]{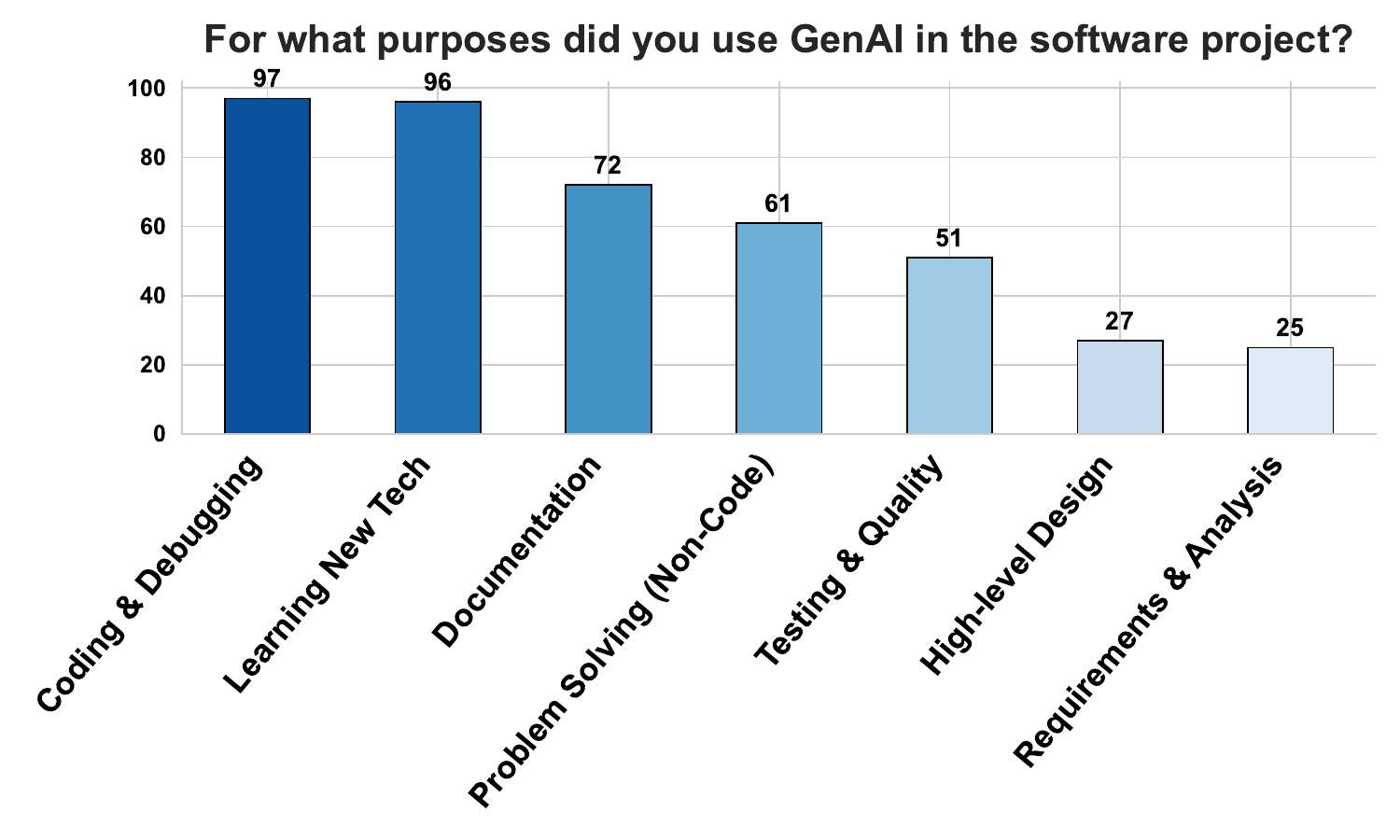}
    \caption{Distribution of use cases for which 150 students used GenAI as part of their software project. Multiple selections were possible.}
    \label{fig:usage}
\end{figure}

\autoref{fig:usage} shows the areas in which students reported using GenAI during the SWP. We predefined seven possible areas of application, and students could additionally specify further use cases in a free-text field. Because the course covers an end-to-end software development process, from requirements elicitation to final testing and presentation, students could select multiple areas of GenAI use.

Students reported using GenAI across all predefined areas. The most frequent uses were code generation or debugging (97) and learning new technologies (e.g., new frameworks or programming languages) (96). In contrast, GenAI was used far less often for high-level design tasks (architecture or technology suggestions) (27) and for requirements gathering or analysis (e.g., researching domain context) (25). Thirteen students provided additional free-text use cases, which mostly fit into predefined categories, but students considered them worth mentioning as ``other''. These included: (1) coding tasks (identifying errors without generating new code, creating an MCP server for GenAI, resolving frontend issues, generating configuration files, designing functions and classes for consistency, generating dependency lists); (2) learning new technologies (installation instructions, legal considerations); (3) preparing the final presentation (poster design ideas); (4) testing (test generation); and (5) documentation (text inspiration). Two students indicated that they used GenAI for either all or none of the listed use cases.

For questions regarding students' assessment of their own programming skills, frequency of GenAI use during the software project, frequency of use of a human-in-the-loop or GenAI-in-the-loop process, general attitude towards the use of GenAI, and assessment of the advantages of GenAI, we performed a correlation analysis to statistically verify correlations and test the hypothesis from \autoref{tab:null_hypotheses}. \autoref{tab:korrelationen} shows the results of this analysis. 
We found a weak negative correlation~\cite{AliAbdAlHameed.2022} ($r_s$ = -.169, $N$ = 150, $p$ = .039) between the self-assessed programming skills of a student and the frequency with which they used GenAI in the context of the software project. These results suggest that students with higher programming skills tend to use GenAI less frequently.
The frequency of GenAI use by a student and their opinion on the benefits of the usage, on the other hand, showed an average positive correlation~\cite{AliAbdAlHameed.2022} ($r_s$ = .517, $N$ = 150, $p$ < .001). Students who can see the benefits of the use of GenAI, therefore, seem to use it more frequently or vice versa. 

\begin{table}[htbp]
    \centering
    \caption{Results of the Spearman rank correlation analysis of the responses from $N = 150$ students}
    \label{tab:korrelationen}
    \begin{tabular}{c p{5cm} c c}
        \toprule
        \textbf{$H_0$} & \textbf{Variables} & \textbf{$r_s$} & \textbf{$p$} \\
        \midrule
        \multirow{2}{*}{$H1_0$} & Self-Assessed Programming Skills \&  & \multirow{2}{*}{-.169} & \multirow{2}{*}{= .039} \\
        & GenAI Usage Frequency & & \\
        \addlinespace
        \multirow{2}{*}{$H2_0$} & GenAI Usage Frequency \& & \multirow{2}{*}{.517} & \multirow{2}{*}{< .001} \\
        & Perceived Benefits of GenAI & & \\
        \addlinespace
        \multirow{2}{*}{$H3_0$} & Attitude Towards GenAI \& & \multirow{2}{*}{.444} & \multirow{2}{*}{< .001} \\
        & GenAI-in-the-Loop & & \\
        \addlinespace
        \multirow{2}{*}{$H4_0$} & Attitude Towards GenAI \& & \multirow{2}{*}{.425} & \multirow{2}{*}{< .001} \\
        & Human-in-the-Loop & & \\
        \addlinespace
        \multirow{2}{*}{$H5_0$} & GenAI-in-the-Loop \& & \multirow{2}{*}{.406} & \multirow{2}{*}{< .001} \\
        & Human-in-the-Loop & & \\
        \bottomrule
    \end{tabular}
    \vspace{1ex}
    \raggedright
    \small \textit{Note: }  $r_s$ = Spearman's rank correlation coefficient
\end{table}

The students' attitude towards the use of GenAI and the frequency of their use of a human-in-the-loop or GenAI-in-the-loop approach were analyzed separately. For both approaches, we found a weak positive correlation~\cite{AliAbdAlHameed.2022} (human-in-the-loop: $r_s$ = .425, $N$ = 150, $p$ < .001, GenAI-in-the-loop: $r_s$ = .444, $N$ = 150, $p$ < .001). A positive attitude towards the use of GenAI, therefore, seems to encourage the use of both the GenAI-in-the-loop and the human-in-the-loop approach. 
Finally, we examined the relationship between the use of human-in-the-loop and GenAI-in-the-loop approaches. Here, too, only a weak positive correlation could be established~\cite{AliAbdAlHameed.2022} ($r_s$ = .406, $N$ = 150, $p$ < .001). This result suggests that the use of one of the two methods does not necessarily lead to its opposite being used less frequently.

\begin{figure}[htbp]
    \centering
    \includegraphics[width=0.45\textwidth]{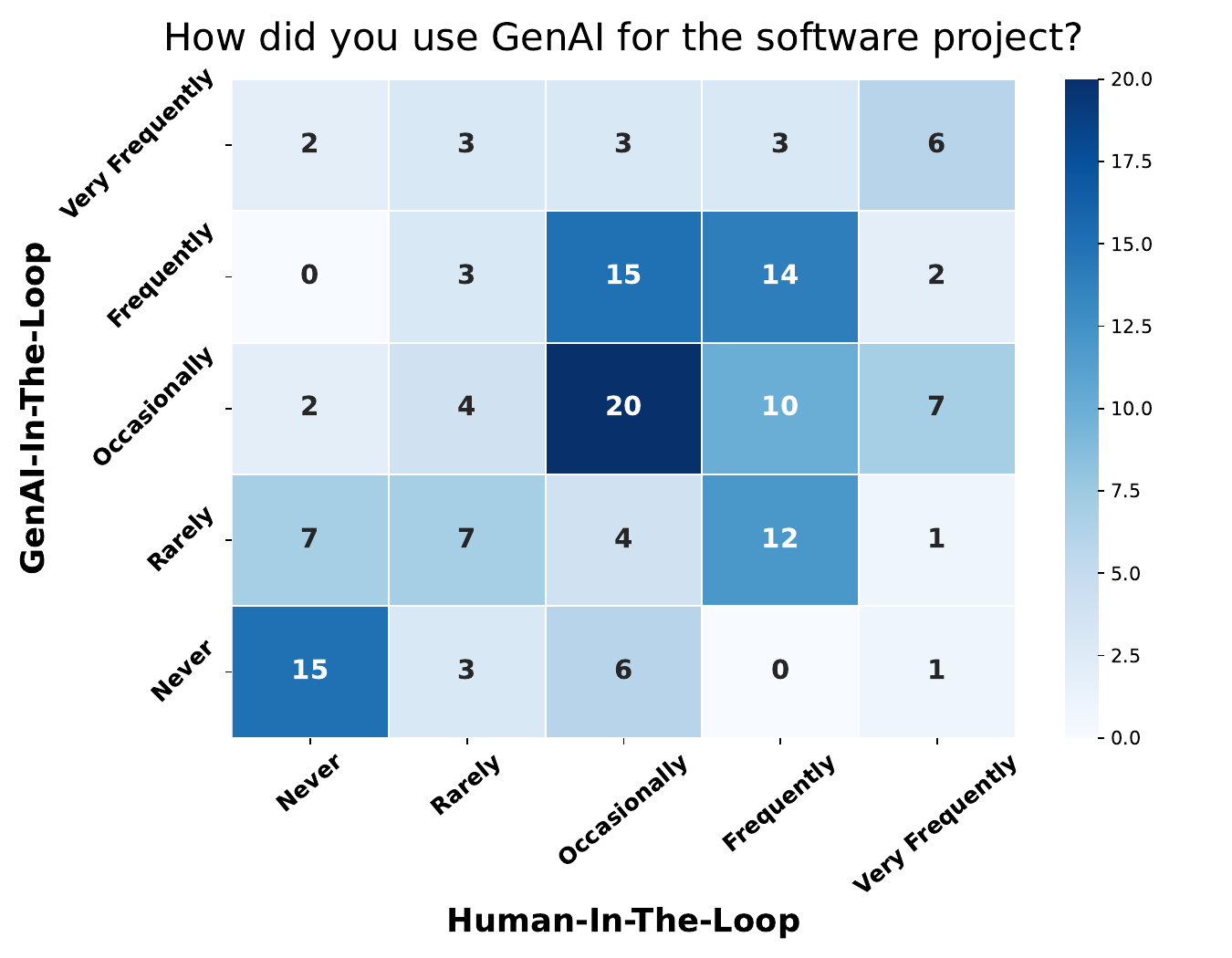}
    \caption{Heatmap of the distribution of Human-In-The-Loop vs. GenAI-In-The-Loop usage by 150 students} 
    \label{fig:heat}
\end{figure}

\autoref{fig:heat} provides an overview of the students' responses on their usage of these two methods. Here, it also becomes apparent that the students apparently either did not use either of the two methods or used both to a similar extent.

\begin{figure}[htbp]
    \centering
    \includegraphics[width=0.45\textwidth]{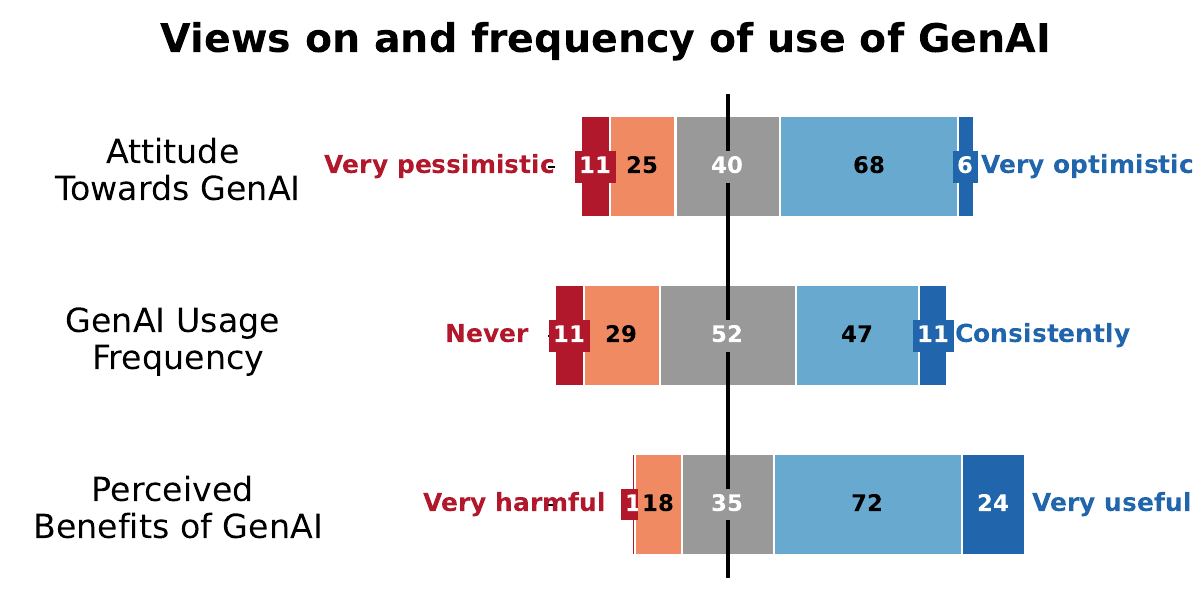}
    \caption{Distribution of the views of students ($N=150$) on GenAI in the context of a RWCP} 
    \label{fig:tornado}
\end{figure}

\autoref{fig:tornado} displays the distribution of the students' attitude towards GenAI, their perceived benefits for students in the context of the software project, and their frequency of use in the SWP. Overall, most students (99) use GenAI at least now and then and have a positive attitude towards its usage. The perceived benefits of GenAI in the context of the SWP are slightly more positive than the students' general attitude towards GenAI.

\begin{table*}[t]
    \centering
    \caption{Distribution of feedback categories and examples adapted from the data set}
    \label{tab:feedbacktable}
    \begin{tabular}{@{}lrrl@{}}
    \toprule
    \textbf{Type of feedback}        & \textbf{\#}  & \textbf{\%}     & \textbf{Example adapted from the data set (translated from German)} \\ \midrule
    All codes combined           & 174          & 100\%           &                  \\ \midrule
    \textbf{Use Case}         & \textbf{57} & \textbf{32.8\%} & \textbf{}        \\
    Learning                    & 19          & 10.9\%          & ``Good for finding an introduction to a completely new field of knowledge.''                 \\
    Debugging                   & 13           & 7.5\%           & ``GenAI is particularly helpful for finding errors.''                 \\
    High-level Design                    & 12           & 6.9\%           & ``AI is very useful when searching for frameworks and how they work.''                 \\
    Trivial Tasks                    & 6           & 3.4\%           & ``Sure, simple things and boilerplate code are fine [...]''                 \\
    Documentation                    & 5           & 2.9\%           & ``However, GenAI is very useful for textual work such as comments, documentation, etc.''                 \\
    User Interface                    & 2           & 1.1\%           & ``Especially in front-end development, it is very helpful [...]''                 \\ \midrule
    \textbf{Directive}     & \textbf{117}  & \textbf{67.2\%} &         \\
    Retain Understanding                    & 35          & 20.1\%          & ``Review the generated code and try to understand it before using it.''                 \\
    Reviewing Outputs                    & 26          & 14.9\%          & ``When GenAI is used to produce code [...], it is essential to check it for mistakes.''                 \\
    Appropriate Degree of Usage                    & 22          & 12.6\%          & ``Useful as an aid, but don't base your entire project on it.''                 \\
    Improving Solutions                    & 10          & 5.7\%          & ``Start by thinking about solutions independently. Then expand/review with AI.''                 \\
    Use Own Solutions                    & 9          & 5.2\%          & ``Do not generate everything from scratch, but work your way forward step by step.''                 \\
    Appropriate Context                    & 8          & 4.6\%          & ``The context window is crucial. [...] \textit{Here, find the error in the entire project} rarely works.''                 \\
    Project-wide Consistency                    & 4          & 2.3\%          & ``Otherwise, you end up with 20 different solutions and ways of writing the same code [...]''                 \\
    Careful Choice of Models                    & 3          & 1.7\%          & ``If you want to use them, you should think about which model is suitable.''                 \\ \bottomrule
    \end{tabular}
\end{table*}

\subsection{Qualitative Student Survey Results}

For the qualitative analysis of the free-text responses, we constructed a coding scheme focusing on two aspects: the use cases for which participants recommended employing GenAI (main category ``Use Case'') and the suggestions for responsible usage (main category ``Directive''). In total, 89 of the 150 participants provided optional free-text recommendations regarding GenAI for future students. Individual statements often contained multiple codes across different categories, resulting in a multi-label coding approach.

\subsubsection{Main Category ``Use Case''}
The ``Use Case'' category consists of six subcategories. Their frequencies and illustrative examples from the survey are reported in Table~\ref{tab:feedbacktable}.

The most frequent subcategory was \emph{Learning} (19), indicating that many participants considered LLMs helpful for acquiring new technologies or becoming familiar with unfamiliar frameworks within their capstone projects. The next most frequent subcategories were \emph{Debugging} (13) and \emph{High-Level Design} (12). Students highlighted the value of LLMs for identifying and correcting errors in their code, as well as for supporting architectural or framework-level decisions and broader development processes. 

The subcategories \emph{Trivial Tasks} (6) and \emph{Documentation} (5) appeared less frequently. These referred to using GenAI to handle simple, repetitive tasks (e.g., boilerplate code) or for assisting with documentation tasks such as writing code comments or refining requirements specifications. Two statements fell under the \emph{User Interface} (2) subcategory, recommending the use of GenAI to support the design and implementation of graphical user interfaces for their software project.

Notably, we did not include a subcategory for code generation. Although many students used GenAI to generate code, they rarely referenced this explicitly in their free-text responses. Instead, several comments—particularly within the ``Directive'' category—implicitly relate to code generation and its associated risks or good practices. Because explicit mentions were too sparse to support a reliable characterization of recommended use cases, we chose not to include code generation as its own ``Use Case'' subcategory.

\subsubsection{Main Category ``Directive''}
The ``Directive'' category comprises eight subcategories whose frequencies and representative examples are summarized in Table~\ref{tab:feedbacktable}.

The most frequent subcategory, \emph{Retain Understanding} (35), concerns students’ emphasis on maintaining comprehension of the project context and preserving learning benefits. Participants stressed the importance of understanding GenAI-generated outputs before integrating them into the project, noting that insufficient understanding may lead to downstream issues. Closely related is the subcategory \emph{Reviewing Outputs} (26), which captures statements urging thorough scrutiny, checking, or testing of GenAI outputs for errors or inconsistencies.

Another common theme was the call for an \emph{Appropriate Degree of Usage} (22). These statements expressed concerns that GenAI may have been overused in some projects and recommended limiting or carefully controlling its use. The subcategories \emph{Improving Solutions} (10) and \emph{Use Own Solutions} (9) address a similar idea: students advised using GenAI primarily to refine or enhance one’s own approaches rather than to generate complete solutions from scratch.

Several participants also highlighted the need to provide \emph{Appropriate Context} (8) when prompting GenAI systems, emphasizing that high-quality outputs require sufficiently detailed and precise contextual information. Without this, results may be unsatisfactory or inconsistent with the existing code base. The remaining two subcategories, \emph{Project-wide Consistency} (4) and \emph{Careful Choice of Models} (3), concern the importance of maintaining consistency across the project when using GenAI and selecting suitable models or tools for different project tasks.

\subsection{Client Survey Results}

Eleven client stakeholders participating in the SWP responded to the client survey. Six indicated that their proposed project explicitly included GenAI-related components, suggesting slightly higher participation from clients already interested in GenAI.

Client attitudes toward GenAI were generally positive: five respondents reported being \emph{very optimistic}, four \emph{rather optimistic}, and two \emph{rather pessimistic}. None rated the high level of student GenAI usage in their project as critical; instead, four viewed it as \emph{harmless}, four as \emph{rather harmless}, and three as \emph{neutral}. For future iterations, only one client stated that GenAI should be used rarely, while three stated students should use it as they see fit and seven would wish for active support to educate and encourage students to use GenAI.

When asked about specific concerns regarding student GenAI usage in this or future SWP iterations, only one client indicated no concerns. Most respondents selected multiple concerns. Table~\ref{tab:client_concerns} summarizes all stated concerns, sorted by frequency. \textit{Lack of student understanding} refers specifically to an inability to converse effectively in client meetings. \textit{Insufficient GenAI training provided} was a specified concern by one client regarding the university's role in students' AI competence.

\begin{table}[htbp]
\centering
\caption{Client concerns regarding student GenAI usage from $N = 11$ clients.}
\label{tab:client_concerns}
\begin{tabular}{l c}
\toprule
\textbf{Concern} & \textbf{Count} \\
\midrule
Lack of student understanding & 7 \\
Insufficient quality & 4 \\
Data protection risks & 4 \\
Transparency / accountability issues & 4 \\
Functional correctness & 3 \\
Intellectual property risks & 2 \\
No concerns & 1 \\
Insufficient GenAI training provided & 1 \\
\bottomrule
\end{tabular}
\end{table}

\section{Discussion}
The findings of our study highlight clear usage patterns, perceived benefits and risks, and a notable convergence between student and client expectations regarding responsible GenAI use. This section discusses these results, acknowledges threats to validity, and outlines implications for software engineering education and future capstone design.

\subsection{Interpretation of Results}
\subsubsection{Use Patterns and Emerging Practices}

Students made substantial use of GenAI throughout the SWP, with 86~\% of students using it for one purpose or another. The most common applications were code generation and debugging, learning unfamiliar technologies, and documentation support. Our proposed taxonomy of use cases was satisfactory in covering all reported use cases, however GenAI was used more than we expected for most of them. Applications recommended by students in qualitative feedback roughly mirror that frequency, suggesting convergence toward early norms for “appropriate” GenAI use in real-world project settings. The only exception to this is ``High-level Design'', which seemed to have been recommended by its users at a relatively higher rate, suggesting this being an underappreciated use case.

We observed a positive correlation between GenAI usage frequency and perceived usefulness. This indicates that students who engage more deeply with GenAI tend to not regret it. One interpretation is that heavy users may discover more effective usage patterns, whereas students with negative predispositions may not use GenAI often enough to realize its benefits. Once students begin using GenAI, however, their usage patterns become pragmatic: general attitudes toward GenAI did not strongly influence whether students preferred AITL or HITL workflows. Instead, workflow choice was driven primarily by overall usage intensity.

The qualitative feedback revealed that many students have a strong awareness of GenAI’s limitations, with almost 60\% of students submitting optional free-text responses. Students frequently emphasized risks such as incorrect or inconsistent code, hallucinations, and recommended practices centered on verification, critical evaluation, and maintaining personal understanding. Two recommendation themes were particularly dominant: using GenAI for learning or understanding technologies, and ensuring that GenAI usage does not compromise independent understanding. These themes suggest that students behave responsibly and reflectively, despite unrestricted access to GenAI tools. 

Nonetheless, this responsible stance may reflect a vocal minority of particularly conscientious respondents. Not all students demonstrated the same level of awareness regarding verification, consistency, or appropriate model selection. For example, most students relied heavily on general-purpose chatbots rather than domain-specific models or integrated tools, contradicting their own recommendations about providing context, choosing appropriate models, and ensuring consistency. This gap suggests that students would benefit from targeted instruction on GenAI tool literacy, including model and tool selection and effective prompting strategies, including output specificity.

\subsubsection{Team Dynamics and the Need for Structured AI Governance}

Students used both AITL and HITL workflows to a similar extent, but qualitative responses revealed that they consider it important to apply these workflows strategically: generating solutions is favored for exploratory, routine, or low-risk tasks, while modifying own solutions is preferred when correctness or reasoning is critical. This emerging differentiation reflects students’ collective awareness of risk and confidence limitations. Recommended usage patterns depending on the project task may be a valuable guideline for students in future iterations. 

Large teams (9--10 students), as in this SWP iteration, introduce additional challenges. GenAI amplifies coordination complexity: different team members may use different AI tools, prompting styles, or models, leading to inconsistent terminology, design drift, or duplicated misunderstandings. Students themselves noted the importance of maintaining team-wide consistency when using GenAI. These findings indicate that GenAI is not merely an individual-support tool but affects collaborative processes and shared understanding between students in complex ways.

Given this variability, one avenue for future course design is the introduction of an additional \emph{AI coordinator} role within each team. Similar to existing roles such as quality assurance lead or roll-out manager, an AI coordinator could oversee consistent tool usage, model selection, responsible-use guidelines, and verification practices across the team. This would help ensure that high-quality GenAI practices observed in a subset of students are disseminated across the group, allowing less experienced or confident users to benefit from emerging team standards.

\subsubsection{Client Perspective}

The client survey revealed a positive stance toward GenAI, however a response bias is evident: over half of responding clients had proposed GenAI-related projects, while only one-third of all projects included GenAI components. This suggests slightly higher engagement among GenAI-interested clients.

The most frequently raised concern was that students might become overly dependent on GenAI and struggle to explain, discuss, or justify decisions during meetings. This aligns strongly with the students’ own emphasis on “retaining understanding,” indicating a shared norm across stakeholders. Both groups implicitly recognize a boundary between appropriate AI support and the need for independent knowledge and reasoning in professional communication.

Other concerns included insufficient quality, functional correctness, transparency and accountability issues, intellectual property risks, and data protection. These concerns mirror real-world SE risks and highlight areas where educators may need to provide explicit safeguards—particularly around data-sensitive usage. Importantly, none of these concerns translate into negative evaluations of the actual student work; rather, they outline expectations for responsible GenAI use in future capstone iterations.

\subsection{Implications for RWCPs}

Overall, the SWP context demonstrates that GenAI is already becoming deeply integrated into student workflows across the SE lifecycle without an active intervention. The findings suggest several implications for future course design:

\begin{itemize}
    \item \textbf{Guidance over restriction:} Responsible students already appear to use GenAI appropriately, and clients support its use. However, both parties acknowledge the risk of over-reliance and losing independent comprehension or lessened learning benefits. Therefore, pedagogical approaches should focus on (1) amplifying the voices of already responsible and effective student users of GenAI within their teams and (2) equipping students with both responsible use guidelines, as well as skills for critical evaluation, verification, and tool selection rather than restricting AI usage.
    \item \textbf{Support for specialized tool literacy:} Students’ relatively low use of specialized GenAI tools indicates a need for explicit instruction on matching AI capabilities to task contexts. However, the very low adoption of the university's own LLM instance LUHKI suggest that getting students to use technologies other than their usual chatbot may be challenging.
    \item \textbf{Team-level GenAI governance:} Introducing an AI coordinator role may help ensure responsible, consistent, and effective practices across large teams and mitigate risks related to heterogeneous tool use and outputs.
    \item \textbf{Preserving authenticity:} Since students perceive the use of GenAI in the SWP as more useful than their general attitude toward the technology would predict, RWCPs offer a uniquely valuable setting for developing authentic AI-supported SE competencies.
\end{itemize}

Together, these points highlight that GenAI is neither uniformly beneficial nor inherently harmful. Instead, its value depends on how well students are prepared to use it responsibly and reflectively in an authentic, results-oriented software development context. The baseline established by this study can guide future pedagogical interventions, inform course design, and help shape emerging best practices for GenAI in software engineering education.

\subsection{Threats to Validity}

We structure the discussion of threats to validity according to the categories proposed by Wohlin et al.~\cite{wohlin2012experimentation}

\subsubsection{Construct Validity}
Our study relies entirely on self-reported data collected at the end of the SWP. This introduces risks related to social desirability bias, particularly because students may wish to avoid creating an impression that GenAI should be restricted for future cohorts. Recall bias is possible as students were asked to retrospectively assess their GenAI usage over a four-month period. Furthermore, since data were collected only at the end of the course, the measured attitudes may already have been shaped by the SWP experience itself, rather than representing stable pre-existing views. Finally, the survey captures perceived usefulness and practices but does not measure the objective effectiveness or quality of GenAI-assisted work, which is difficult to assess given the strong heterogeneity of projects.

\subsubsection{Internal Validity}
The SWP’s substantial variability in project domains, technologies, team compositions, and tutor styles introduces many potential confounding factors that cannot be controlled for in our analysis. Two-thirds of students lacked prior professional software development experience, which may influence GenAI usage patterns and perceived benefits. The clustering of students within teams and projects may also affect results, as team-level dynamics can shape how GenAI is adopted and discussed. These factors limit causal interpretations.

\subsubsection{External Validity}
The study was conducted in a single large-scale RWCP at one institution and under conditions of high ecological authenticity but unusually large team sizes~\cite{tenhunen2023systematic}. This may limit the generalizability of findings to other institutions or capstone formats. Some of the reported use cases (e.g., learning new technologies, suggesting frameworks) are specific to educational settings and may generalize less directly to professional software engineering. The sample of client responses also reflects mild self-selection bias: clients already interested in GenAI were more likely to participate, limiting generalizability of client perspectives.

\subsubsection{Conclusion Validity}
The quantitative analysis is correlational and based on moderate effect sizes; thus, observed relationships should not be interpreted causally. Although qualitative data were coded systematically with substantial inter-rater agreement, interpretation bias cannot be excluded. Survey response-rate (150 of 178 students) introduces the possibility that students with different attitudes or usage patterns are underrepresented. Finally, the small client sample (11 clients from 15 organizations) reduces confidence in conclusions drawn about client expectations and concerns.

\section{Conclusion and Outlook}

This paper presented a large-scale baseline study of how undergraduate software engineering students autonomously use generative AI in Real-World Capstone Projects (RWCPs). Based on mixed-method survey data from 150 students and 11 client stakeholders, we characterized current GenAI usage patterns, identified recommended use cases and responsible-use directives, and documented client expectations and concerns. Our results show that GenAI is already deeply embedded in students’ development workflows across the SE lifecycle, with code-related tasks, learning new technologies, and documentation emerging as dominant use cases. Students generally express a reflective and cautious stance toward GenAI, emphasizing verification, maintaining independent understanding, and avoiding over-reliance. Clients supported student GenAI use but expressed expectations regarding student comprehension in client meetings, quality assurance, and data protection.

These findings establish an empirical baseline of self-determined GenAI appropriation in RWCPs and highlight several directions for pedagogical support. First, guidance rather than restriction appears most appropriate, and responsible students already demonstrate effective practices that can serve as models for their peers. Second, GenAI tool literacy may be underdeveloped, especially regarding specialized tools and model selection. Third, RWCPs would benefit from team-level AI governance structures (such as an \emph{AI coordinator} role) to reduce inconsistency, support responsible practices, and disseminate effective workflows. Finally, students’ generally positive perception of GenAI's usefulness in the SWP suggests that authentic project settings are uniquely suited to cultivating informed and reflective AI use.

Looking forward, this baseline enables an iterative improvement process for future RWCP iterations. We plan to translate the identified best practices into explicit responsible-use guidelines, provide targeted learning resources to strengthen students’ AI literacy, and introduce structural elements—such as an AI coordinator—to support consistent and reflective GenAI use within teams. Future cohorts will allow us to reassess emerging use cases, evaluate the impact of these interventions, and explore opportunities for dedicated tool support in areas where GenAI shows clear promise. Through this iterative approach, we aim to develop a sustainable model for integrating GenAI into software engineering education while preserving the authenticity and learning value of RWCPs.

\section{Data Availability Statement}
Full survey questionnaires, coding procedure details, survey results, data visualization, and statistical analyses can be found in our supplementary material~\cite{mircea_2026_18836366}.

\begin{acks}
 Several LLMs were used for AI-assisted copy editing to improve language, formatting, and structure of this paper.
\end{acks}

\bibliographystyle{ACM-Reference-Format}
\bibliography{references}
\end{document}